\documentclass[twocolumn,letterpaper,aps,prd,longbibliography,superscriptaddress,showpacs,nofootinbib,floatfix]{revtex4-2}
\usepackage{graphicx}	% Include figure files
\usepackage{xspace}	% Include xspace

\newcommand{\pt}{\mbox{$p_T$}\xspace}

\newcommand{\sqsn}{\mbox{$\sqrt{s_{_{NN}}}$}\xspace}

\newcommand{\pp}{\mbox{$p$$+$$p$}\xspace}
\newcommand{\pA}{\mbox{$p$$+$$A$}\xspace}
\newcommand{\pau}{\mbox{$p$$+$Au}\xspace}
\newcommand{\pal}{\mbox{$p$$+$Al}\xspace}

\newcommand{\piz}{\mbox{$\pi^{0}$}\xspace}
\newcommand{\ptp}{\mbox{$p^\uparrow$$+$$p$}\xspace}
\newcommand{\pta}{\mbox{$p^\uparrow$$+$$A$}\xspace}
\newcommand{\ptau}{\mbox{$p^\uparrow$$+$Au}\xspace}
\newcommand{\ptal}{\mbox{$p^\uparrow$$+$Al}\xspace}
\newcommand{\gevc}{{\rm GeV}/c}

\def\figure#1{Figure~\ref{#1}}
\def\fig#1{Fig.~\ref{#1}}
\def\eq#1{Eq.~(\ref{#1})}

\begin{document}

%%%%% NOTE:  Use the above macros in text and captions, but not in
%%%%%        the title or abstract.  Those two elements must stand
%%%%%        alone and not be dependent on the macros defined here.


\title{Transverse single-spin asymmetry of midrapidity $\pi^{0}$ and $\eta$
mesons in $p$$+$Au and $p$$+$Al collisions at $\sqrt{s_{_{NN}}}=$ 200 GeV}

\newcommand{\abilene}{Abilene Christian University, Abilene, Texas 79699, USA}
\newcommand{\augie}{Department of Physics, Augustana University, Sioux Falls, South Dakota 57197, USA}
\newcommand{\banaras}{Department of Physics, Banaras Hindu University, Varanasi 221005, India}
\newcommand{\barc}{Bhabha Atomic Research Centre, Bombay 400 085, India}
\newcommand{\baruch}{Baruch College, City University of New York, New York, New York, 10010 USA}
\newcommand{\bnlcoll}{Collider-Accelerator Department, Brookhaven National Laboratory, Upton, New York 11973-5000, USA}
\newcommand{\bnlphys}{Physics Department, Brookhaven National Laboratory, Upton, New York 11973-5000, USA}
\newcommand{\caucr}{University of California-Riverside, Riverside, California 92521, USA}
\newcommand{\charlesczech}{Charles University, Faculty of Mathematics and Physics, 180 00 Troja, Prague, Czech Republic}
\newcommand{\cns}{Center for Nuclear Study, Graduate School of Science, University of Tokyo, 7-3-1 Hongo, Bunkyo, Tokyo 113-0033, Japan}
\newcommand{\colorado}{University of Colorado, Boulder, Colorado 80309, USA}
\newcommand{\columbia}{Columbia University, New York, New York 10027 and Nevis Laboratories, Irvington, New York 10533, USA}
\newcommand{\czechtech}{Czech Technical University, Zikova 4, 166 36 Prague 6, Czech Republic}
\newcommand{\debrecen}{Debrecen University, H-4010 Debrecen, Egyetem t{\'e}r 1, Hungary}
\newcommand{\elte}{ELTE, E{\"o}tv{\"o}s Lor{\'a}nd University, H-1117 Budapest, P{\'a}zm{\'a}ny P.~s.~1/A, Hungary}
\newcommand{\ewha}{Ewha Womans University, Seoul 120-750, Korea}
\newcommand{\famu}{Florida A\&M University, Tallahassee, FL 32307, USA}
\newcommand{\fsu}{Florida State University, Tallahassee, Florida 32306, USA}
\newcommand{\gsu}{Georgia State University, Atlanta, Georgia 30303, USA}
\newcommand{\hiroshima}{Physics Program and International Institute for Sustainability with Knotted Chiral Meta Matter (SKCM2), Hiroshima University, Higashi-Hiroshima, Hiroshima 739-8526, Japan}
\newcommand{\howard}{Department of Physics and Astronomy, Howard University, Washington, DC 20059, USA}
\newcommand{\ihepprot}{IHEP Protvino, State Research Center of Russian Federation, Institute for High Energy Physics, Protvino, 142281, Russia}
\newcommand{\illuiuc}{University of Illinois at Urbana-Champaign, Urbana, Illinois 61801, USA}
\newcommand{\inrras}{Institute for Nuclear Research of the Russian Academy of Sciences, prospekt 60-letiya Oktyabrya 7a, Moscow 117312, Russia}
\newcommand{\instpasczech}{Institute of Physics, Academy of Sciences of the Czech Republic, Na Slovance 2, 182 21 Prague 8, Czech Republic}
\newcommand{\isu}{Iowa State University, Ames, Iowa 50011, USA}
\newcommand{\jaea}{Advanced Science Research Center, Japan Atomic Energy Agency, 2-4 Shirakata Shirane, Tokai-mura, Naka-gun, Ibaraki-ken 319-1195, Japan}
\newcommand{\jeonbuk}{Jeonbuk National University, Jeonju, 54896, Korea}
\newcommand{\kek}{KEK, High Energy Accelerator Research Organization, Tsukuba, Ibaraki 305-0801, Japan}
\newcommand{\korea}{Korea University, Seoul 02841, Korea}
\newcommand{\kurchatov}{National Research Center ``Kurchatov Institute", Moscow, 123098 Russia}
\newcommand{\kyoto}{Kyoto University, Kyoto 606-8502, Japan}
\newcommand{\lawllnl}{Lawrence Livermore National Laboratory, Livermore, California 94550, USA}
\newcommand{\losalamos}{Los Alamos National Laboratory, Los Alamos, New Mexico 87545, USA}
\newcommand{\lund}{Department of Physics, Lund University, Box 118, SE-221 00 Lund, Sweden}
\newcommand{\lyon}{IPNL, CNRS/IN2P3, Univ Lyon, Universit{\'e} Lyon 1, F-69622, Villeurbanne, France}
\newcommand{\maryland}{University of Maryland, College Park, Maryland 20742, USA}
\newcommand{\mass}{Department of Physics, University of Massachusetts, Amherst, Massachusetts 01003-9337, USA}
\newcommand{\mate}{MATE, Laboratory of Femtoscopy, K\'aroly R\'obert Campus, H-3200 Gy\"ongy\"os, M\'atrai\'ut 36, Hungary}
\newcommand{\michigan}{Department of Physics, University of Michigan, Ann Arbor, Michigan 48109-1040, USA}
\newcommand{\miss}{Mississippi State University, Mississippi State, Mississippi 39762, USA}
\newcommand{\muhlenberg}{Muhlenberg College, Allentown, Pennsylvania 18104-5586, USA}
\newcommand{\nara}{Nara Women's University, Kita-uoya Nishi-machi Nara 630-8506, Japan}
\newcommand{\natmephi}{National Research Nuclear University, MEPhI, Moscow Engineering Physics Institute, Moscow, 115409, Russia}
\newcommand{\newmex}{University of New Mexico, Albuquerque, New Mexico 87131, USA}
\newcommand{\nmsu}{New Mexico State University, Las Cruces, New Mexico 88003, USA}
\newcommand{\northcg}{Physics and Astronomy Department, University of North Carolina at Greensboro, Greensboro, North Carolina 27412, USA}
\newcommand{\ohio}{Department of Physics and Astronomy, Ohio University, Athens, Ohio 45701, USA}
\newcommand{\ornl}{Oak Ridge National Laboratory, Oak Ridge, Tennessee 37831, USA}
\newcommand{\orsay}{IPN-Orsay, Univ.~Paris-Sud, CNRS/IN2P3, Universit\'e Paris-Saclay, BP1, F-91406, Orsay, France}
\newcommand{\peking}{Peking University, Beijing 100871, People's Republic of China}
\newcommand{\pnpi}{PNPI, Petersburg Nuclear Physics Institute, Gatchina, Leningrad region, 188300, Russia}
\newcommand{\pusan}{Pusan National University, Pusan 46241, Korea}
\newcommand{\riken}{RIKEN Nishina Center for Accelerator-Based Science, Wako, Saitama 351-0198, Japan}
\newcommand{\rikjrbrc}{RIKEN BNL Research Center, Brookhaven National Laboratory, Upton, New York 11973-5000, USA}
\newcommand{\rikkyo}{Physics Department, Rikkyo University, 3-34-1 Nishi-Ikebukuro, Toshima, Tokyo 171-8501, Japan}
\newcommand{\saispbstu}{Saint Petersburg State Polytechnic University, St.~Petersburg, 195251 Russia}
\newcommand{\seoulnat}{Department of Physics and Astronomy, Seoul National University, Seoul 151-742, Korea}
\newcommand{\stonybrkc}{Chemistry Department, Stony Brook University, SUNY, Stony Brook, New York 11794-3400, USA}
\newcommand{\stonycrkp}{Department of Physics and Astronomy, Stony Brook University, SUNY, Stony Brook, New York 11794-3800, USA}
\newcommand{\tenn}{University of Tennessee, Knoxville, Tennessee 37996, USA}
\newcommand{\texsu}{Texas Southern University, Houston, TX 77004, USA}
\newcommand{\titech}{Department of Physics, Tokyo Institute of Technology, Oh-okayama, Meguro, Tokyo 152-8551, Japan}
\newcommand{\tsukuba}{Tomonaga Center for the History of the Universe, University of Tsukuba, Tsukuba, Ibaraki 305, Japan}
\newcommand{\vandy}{Vanderbilt University, Nashville, Tennessee 37235, USA}
\newcommand{\weizmann}{Weizmann Institute, Rehovot 76100, Israel}
\newcommand{\wigner}{Institute for Particle and Nuclear Physics, Wigner Research Centre for Physics, Hungarian Academy of Sciences (Wigner RCP, RMKI) H-1525 Budapest 114, POBox 49, Budapest, Hungary}
\newcommand{\yonsei}{Yonsei University, IPAP, Seoul 120-749, Korea}
\newcommand{\zagreb}{Department of Physics, Faculty of Science, University of Zagreb, Bijeni\v{c}ka c.~32 HR-10002 Zagreb, Croatia}
\newcommand{\zambia}{Department of Physics, School of Natural Sciences, University of Zambia, Great East Road Campus, Box 32379, Lusaka, Zambia}
\affiliation{\abilene}
\affiliation{\augie}
\affiliation{\banaras}
\affiliation{\barc}
\affiliation{\baruch}
\affiliation{\bnlcoll}
\affiliation{\bnlphys}
\affiliation{\caucr}
\affiliation{\charlesczech}
\affiliation{\cns}
\affiliation{\colorado}
\affiliation{\columbia}
\affiliation{\czechtech}
\affiliation{\debrecen}
\affiliation{\elte}
\affiliation{\ewha}
\affiliation{\famu}
\affiliation{\fsu}
\affiliation{\gsu}
\affiliation{\hiroshima}
\affiliation{\howard}
\affiliation{\ihepprot}
\affiliation{\illuiuc}
\affiliation{\inrras}
\affiliation{\instpasczech}
\affiliation{\isu}
\affiliation{\jaea}
\affiliation{\jeonbuk}
\affiliation{\kek}
\affiliation{\korea}
\affiliation{\kurchatov}
\affiliation{\kyoto}
\affiliation{\lawllnl}
\affiliation{\losalamos}
\affiliation{\lund}
\affiliation{\lyon}
\affiliation{\maryland}
\affiliation{\mass}
\affiliation{\mate}
\affiliation{\michigan}
\affiliation{\miss}
\affiliation{\muhlenberg}
\affiliation{\nara}
\affiliation{\natmephi}
\affiliation{\newmex}
\affiliation{\nmsu}
\affiliation{\northcg}
\affiliation{\ohio}
\affiliation{\ornl}
\affiliation{\orsay}
\affiliation{\peking}
\affiliation{\pnpi}
\affiliation{\pusan}
\affiliation{\riken}
\affiliation{\rikjrbrc}
\affiliation{\rikkyo}
\affiliation{\saispbstu}
\affiliation{\seoulnat}
\affiliation{\stonybrkc}
\affiliation{\stonycrkp}
\affiliation{\tenn}
\affiliation{\texsu}
\affiliation{\titech}
\affiliation{\tsukuba}
\affiliation{\vandy}
\affiliation{\weizmann}
\affiliation{\wigner}
\affiliation{\yonsei}
\affiliation{\zagreb}
\affiliation{\zambia}
\author{N.J.~Abdulameer} \affiliation{\debrecen}
\author{U.~Acharya} \affiliation{\gsu} 
\author{C.~Aidala} \affiliation{\michigan} 
\author{Y.~Akiba} \email[PHENIX Spokesperson: ]{akiba@rcf.rhic.bnl.gov} \affiliation{\riken} \affiliation{\rikjrbrc} 
\author{M.~Alfred} \affiliation{\howard} 
\author{V.~Andrieux} \affiliation{\michigan} 
\author{N.~Apadula} \affiliation{\isu} 
\author{H.~Asano} \affiliation{\kyoto} \affiliation{\riken} 
\author{B.~Azmoun} \affiliation{\bnlphys} 
\author{V.~Babintsev} \affiliation{\ihepprot} 
% ...
\collaboration{PHENIX Collaboration}  \noaffiliation

\date{\today}

%------------------------------------------------------------------------------|

\begin{abstract}

%\linenumbers

Presented are the first measurements of the transverse 
single-spin asymmetries ($A_N$) for neutral pions and eta 
mesons in $p$$+$Au and $p$$+$Al collisions at 
$\sqrt{s_{_{NN}}}=200$ GeV in the pseudorapidity range 
$|\eta|<$0.35 with the PHENIX detector at the Relativistic 
Heavy Ion Collider.  The asymmetries are consistent with zero, 
similar to those for midrapidity neutral pions and eta mesons 
produced in $p$$+$$p$ collisions. These measurements show no 
evidence of additional effects that could potentially arise 
from the more complex partonic environment present in 
proton-nucleus collisions.

\end{abstract}

%\maketitle must follow title, authors, abstract, and \keywords

\maketitle

\section{Introduction}

Transverse single-spin asymmetries (TSSAs) in particle production for 
hadronic collisions involving a transversely polarized proton result 
from nonperturbative spin-momentum correlations in the proton and/or the 
process of hadronization~\cite{TSSA_review}.  For recent discussions of 
TSSAs measured in polarized \pp collisions at the Relativistic Heavy Ion 
Collider (RHIC) and the possible mechanisms contributing to them, see 
Refs.~\cite{PPG246,PPG247,PPG235,PPG234,PPG238,STAR:2022hqg,STAR:2020nnl,STAR:2017wsi,STAR:2017akg}.

In hadronic collisions involving a nucleus, the underlying partonic 
origins of the asymmetries could be affected by the presence of more 
complex quantum-chromodynamics environments. For example, relations 
between TSSAs and the physics of small parton momentum fractions have 
been proposed, in particular how comparisons of asymmetries measured in 
\ptp and \pta collisions for forward hadron production could be used to 
probe gluon saturation effects in the nucleus~\cite{Kang:2011ni}. 
Further theoretical works have explored these 
ideas~\cite{Kovchegov:2012ga,Schafer:2014xpa,Schafer:2014zea,Kovchegov:2015zha,Zhou:2015ima,Hatta:2016wjz,Hatta:2016khv,Zhou:2017sdx,Benic:2018amn,Kovchegov:2020kxg,Benic:2022qzv}. 
At RHIC, TSSA measurements for proton-nucleus collisions have been 
performed for forward charged hadrons~\cite{PPG215, PPG226} and forward 
$J/\psi$ mesons~\cite{PPG211} by PHENIX, and for forward $\pi^0$ 
production by STAR~\cite{STAR:2020grs}, revealing some nuclear 
dependencies that remain to be understood in detail. PHENIX has 
additionally measured the TSSAs for far forward neutron production, with 
the observed nuclear dependence of the asymmetries understood to be due 
to the interplay of hadronic and electromagnetic interactions in 
ultra-peripheral collisions~\cite{PPG203,PPG244}. No experimental 
measurements exist and very little theoretical work has been done to 
explore possible nuclear effects for midrapidity TSSA observables, which 
can only be studied at RHIC.
 
\vspace{0.5cm}

\section{Analysis}

%-------------------------------------------- Fig_1
%\begin{minipage}{0.48\linewidth}
\begin{figure*}[tbh]
  \includegraphics[width=0.47\linewidth]{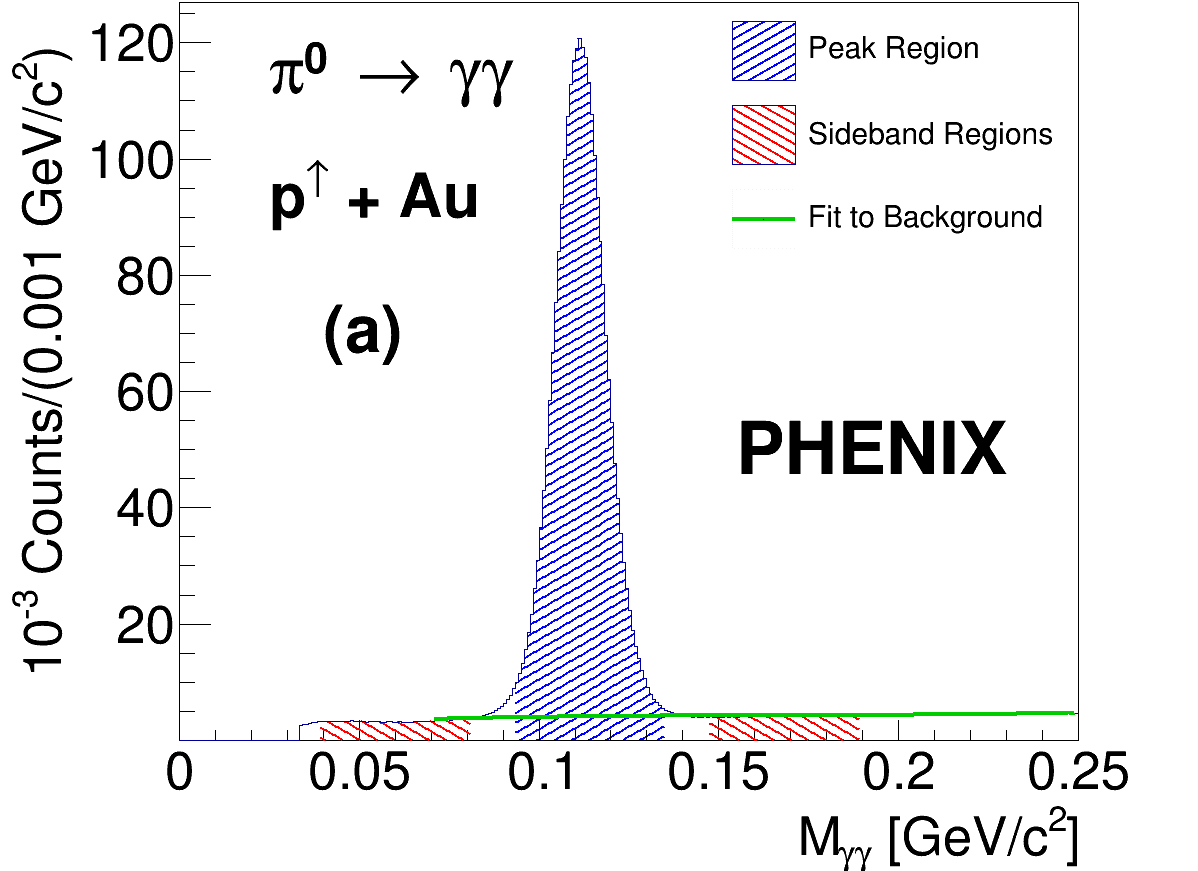}
  \includegraphics[width=0.47\linewidth]{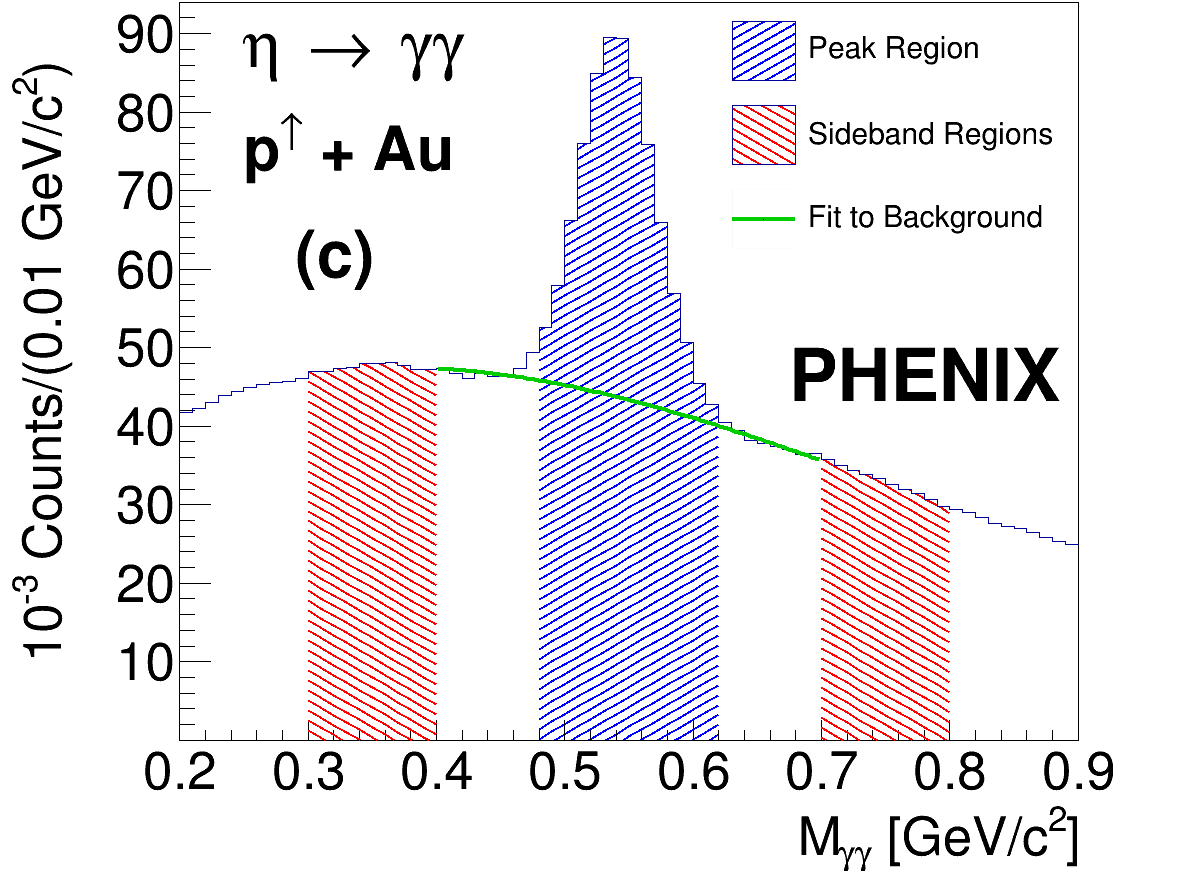}
  \includegraphics[width=0.47\linewidth]{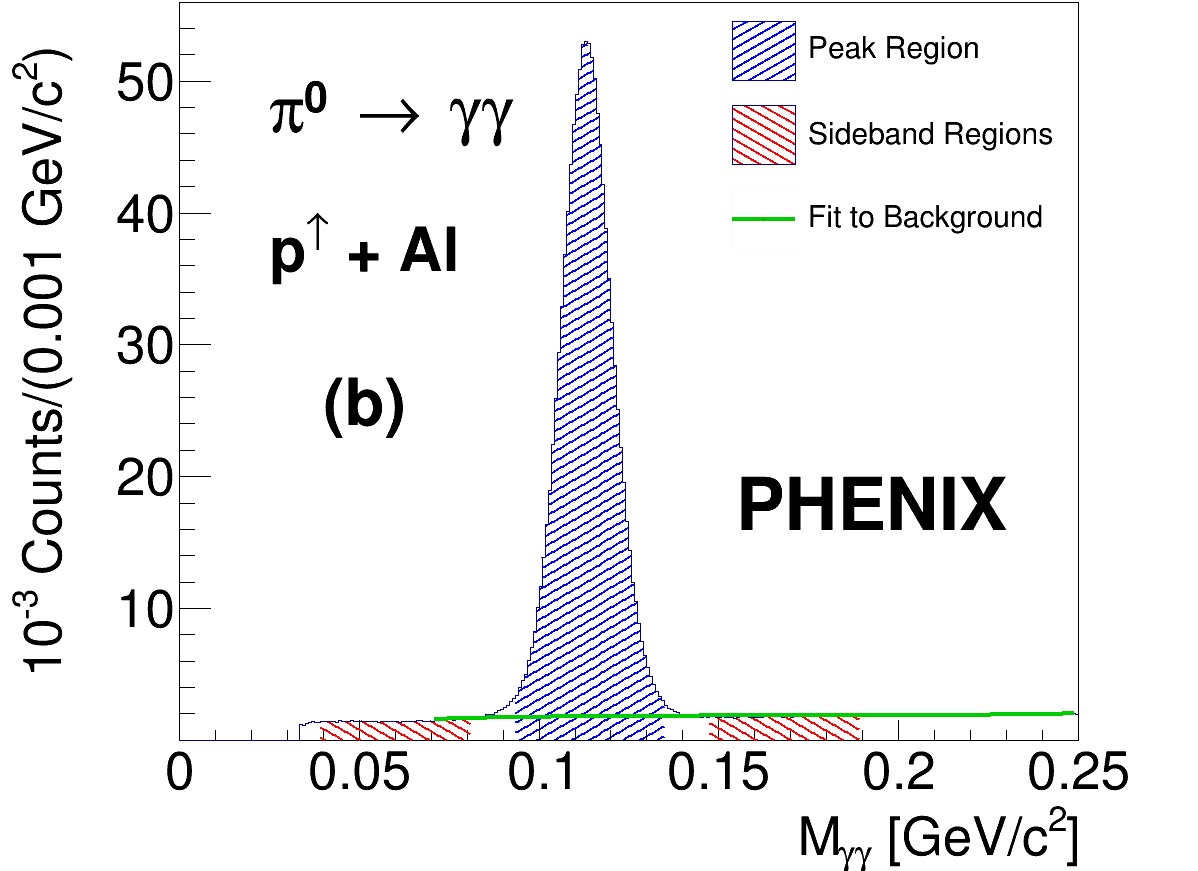} 
  \includegraphics[width=0.47\linewidth]{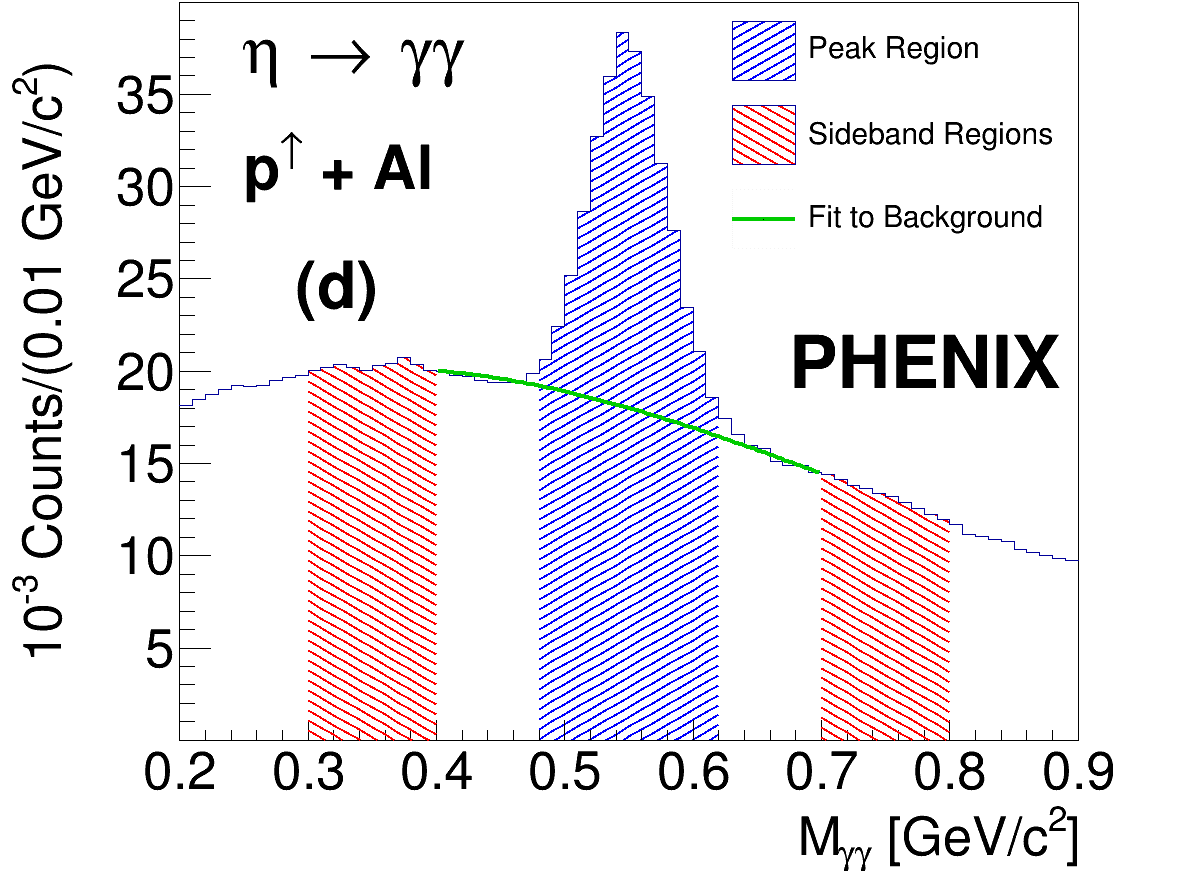} 
  \caption{Invariant mass distributions around the 
$\pi^{0} \rightarrow \gamma\gamma$ peak in 
(a) \ptau collisions and 
(b) \ptal collisions 
and around the $\eta \rightarrow \gamma\gamma$ peak in 
(c) \ptau collisions and 
(d) \ptal collisions for photon pairs within 
$4<p_T~[{\rm GeV}/c]<5$ in the west central-arm spectrometer. The 
[blue] leftward-hatched regions are the signal peaks, used for quantifying 
yields for the $A_{N}$ calculations, the [red] rightward-hatched regions 
are the side bands, used to quantify yields for the $A_{N}^{BG}$ 
calculations, and the [green] solid curves correspond to fits to the 
combinatorial background, used in calculating the background fractions.
}
  \label{fig:invmass}
\end{figure*}

This brief article reports the first measurement of the TSSAs of neutral 
pions and eta mesons in proton-Gold (\ptau) and proton-Aluminum (\ptal) 
collisions at \sqsn=~200 GeV at midrapidity ($|\eta|<$0.35). The 
data were taken in 2015 at RHIC and total integrated luminosities of 
approximately 202 and 690 nb$^{-1}$, respectively, were collected.

Measurements were performed with collisions of a vertically polarized 
proton beam on an ion beam (Au or Al). The proton or ion bunches are 
separated by $106$~ns in the RHIC rings. Each polarized proton bunch is 
assigned a polarization direction, either up or down, so that 
measurements with both spin directions can be performed nearly 
simultaneously. This significantly reduces any possible systematic 
uncertainties related to the detector performance with time. The average 
proton beam polarization was $0.60$ and $0.57$ in \ptau and \ptal 
collisions respectively~\cite{RHIC_polarimetry}, with a relative 
uncertainty of $3\%$ due to uncertainty in polarization normalization.

The data analysis procedure follows almost exactly from the recent TSSA 
measurement for $\pi^{0}$ and $\eta$ mesons in $\sqrt{s} = 200$ GeV polarized 
$p$$+$$p$ collisions~\cite{PPG234}, with the distinction that only the proton 
beam is polarized in \pta collisions. Only events with a 
collision z-vertex within $\pm 30~{\rm cm}$ from the nominal collision 
point were selected. The collision or minimum-bias trigger, as well 
as vertex position, were determined by two beam-beam counters (BBC) 
located at $\pm 144~{\rm cm}$ from the nominal collision point along the 
beam line, and covering the pseudorapidity range $3.1<|\eta|<3.9$ with 
full azimuthal coverage.

Neutral pions and eta mesons were reconstructed through their two-photon 
decay in the electromagnetic calorimeters (EMCal) of PHENIX. The EMCal 
is located in two nearly back-to-back central arm spectrometers (west 
and east), each covering $\Delta\phi=\pi/2$ in azimuth and $\pm0.35$ in 
pseudorapidity. The EMCal comprises two types of calorimeters, six 
sectors of sampling lead-scintillator (PbSc) calorimeter and 2 sectors 
of $\check{\rm C}$erenkov lead-glass (PbGl) 
calorimeter~\cite{Aphecetche:2003zr}.  The two calorimeter systems have 
different granularity 
($\Delta\phi \times \Delta \eta =0.011 \times 0.011$ 
in PbSc and $0.008 \times 0.008$ in PbGl) and also have a 
different response to charged hadrons, which provides important 
systematic cross checks for the measurement.

The PHENIX EMCal was also used to generate a high-\pt photon trigger to 
tag events with a high-energy cluster in the EMCal. The high-\pt photon 
trigger (with an energy threshold of 1.5 GeV) in coincidence with a 
minimum-bias trigger that requires charged particles in both BBC detectors was used to collect the \piz and $\eta$ statistics 
in this analysis. The efficiency of such a trigger for \piz's increased 
from 20\% at $\pt=3~\gevc$ to 90\% at $\pt>6~\gevc$, with the plateaued 
efficiency level defined by the acceptance of the live trigger tiles.

Photons were identified in the EMCal by placing selection criteria on 
the shower profile and time-of-flight (TOF) with $|{\rm TOF}| < 5$ 
ns, and with a minimum energy selection of 0.5 GeV to reduce the 
contribution from electronic noise in the EMCal, and combinatorial 
background in $\pi^{0}$ and $\eta$ reconstruction.  A charged track veto 
was also implemented to eliminate clusters that are geometrically 
associated with a track and to suppress the background from 
electrons and charged hadrons. Photon pairs were reconstructed by 
finding a high-$p_{T}$ trigger photon, and pairing it with another 
photon from the same event and spectrometer arm. Photon pairs passing an 
energy-asymmetry requirement: $\alpha = |E_1 - E_2|/(E_1+E_2)<0.8$ were 
selected for further analysis. 

\figure{fig:invmass} shows the two-photon invariant mass distributions 
around the \piz and $\eta$ peaks for photon pairs within $4<p_T<5$ 
GeV/$c$ in the west central arm spectrometer for \pau and \pal 
collisions.  The \piz and $\eta$ meson yields were defined to be within 
the signal-invariant mass window ([blue] leftward-hatched regions in 
\fig{fig:invmass}) of $\pm$~25 and $\pm$~70 MeV/$c^{2}$ from the \piz 
and $\eta$ mass peaks, respectively, in the two-photon invariant mass 
distribution for each \pt bin. The side-band regions used to approximate 
the combinatorial background under the signal peak ([red] 
rightward-hatched regions in \fig{fig:invmass}) are defined as 47--97 
and 177--227 MeV/$c^{2}$ for the \piz and 300--400 and 700--800 
MeV/c$^{2}$ for the $\eta$ mesons.  The same signal and side-band regions 
were used in a previous PHENIX analysis~\cite{PPG234}. The combinatorial 
background for the two-photon invariant mass spectrum (described by a 
third-order polynomial and shown as the [green] solid lines in 
\fig{fig:invmass}) was used to quantify the fraction of background 
existing under the signal peaks.  For the \piz, this ranged from 14\% 
(13\%) to 6\% (6\%) from the lowest to the highest \pt bins in \pau 
(\pal) collisions, while for the $\eta$, the combinatorial background 
under the signal peak ranged from 79\% (77\%) to 48\% (43\%) in \pau 
(\pal) collisions.

%-------------------------------------------- Fig_2
    \begin{figure*}[tbh]
        \includegraphics[width=0.48\linewidth]{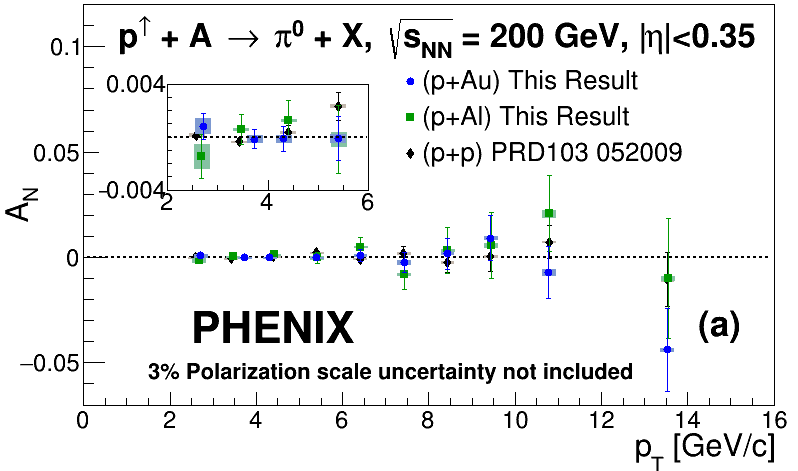}
        \includegraphics[width=0.48\linewidth]{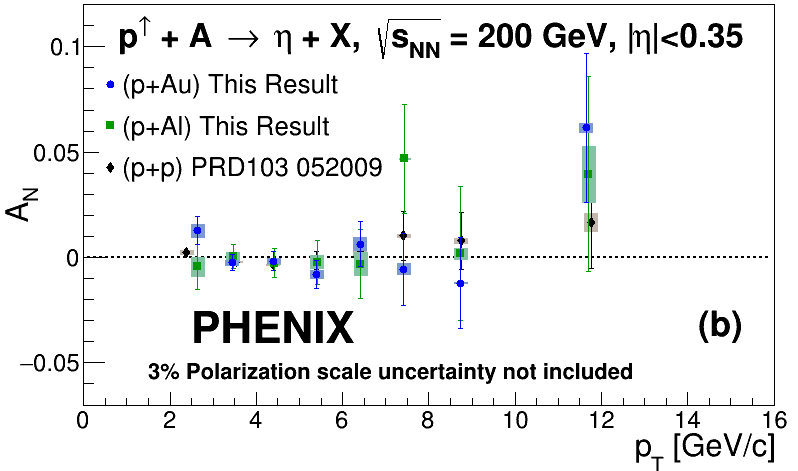}
\caption{Transverse single-spin asymmetry for (a) $\pi^{0}$ and (b) $\eta$ 
mesons in \ptau collisions ([blue] circles), and \ptal collisions ([green] 
squares) from this measurement, shown alongside the same measurement in 
polarized $p$$+$$p$ collisions from Ref.~\cite{PPG234} ([black] diamonds). 
The error bars represent the statistical uncertainty 
($\sigma^{\rm stat}$) while the boxes represent the total systematic 
uncertainty ($\sigma^{\rm syst}$).
}
        \label{fig:ANpT}
    \end{figure*}

%================================================ Table_I
    \begin{table*}[htb]
\caption{Summary of final asymmetries with statistical and
systematic uncertainties for $\pi^{0}$ and $\eta$ mesons in \pta collisions.
$\sigma^{\rm syst}$ corresponds to the systematic uncertainties, displayed
by the shaded boxes in Fig.~\protect\ref{fig:ANpT}.
}
        \begin{ruledtabular}  \begin{tabular}{ccccccc}
Meson & Collisions & $p_{T}$ range [\gevc] & $\langle\pt\rangle [\gevc]$ 
                   & $A_{N}$ & $\sigma^{\rm stat}$ & $\sigma^{\rm syst}$ \\
                \hline
$\pi^{0}$ &  \ptau & 2--3 & 2.71 & 0.000818 & 0.000993 & 0.000569 \\
         &      & 3--4 & 3.73 & -0.000145 & 0.000701 & 0.000286 \\
         &       & 4--5 & 4.31 & -0.000135 & 0.000974 & 0.000257 \\
         &       & 5--6 & 5.4 & -0.00011 & 0.00164 & 0.00028 \\
         &       & 6--7 & 6.41 & 0.00097 & 0.00281 & 0.00024 \\
         &       & 7--8 & 7.42 & -0.00243 & 0.00464 & 0.00109 \\
         &       & 8--9 & 8.43 & 0.00179 & 0.00732 & 0.00055 \\
         &       & 9--10 & 9.44 & 0.0093 & 0.0106 & 0.0005 \\
         &       & 10--12 & 10.8 & -0.0072 & 0.0122 & 0.0014 \\
         &       & 12--20 & 13.5 & -0.0438 & 0.0198 & 0.0008 \\
                \\
$\pi^{0}$ &       \ptal & 2--3 & 2.67 & -0.00147 & 0.00163 & 0.00088 \\
         &       & 3--4 & 3.47 & 0.00056 & 0.00113 & 0.00006 \\
         &       & 4--5 & 4.41 & 0.00126 & 0.00153 & 0.00005 \\
         &       & 5--6 & 5.41 & -0.00018 & 0.00254 & 0.00051 \\
         &       & 6--7 & 6.42 & 0.00500 & 0.00429 & 0.00042 \\
         &       & 7--8 & 7.42 & -0.00809 & 0.00699 & 0.00060 \\
         &       & 8--9 & 8.43 & 0.0035 & 0.0109 & 0.0004 \\
         &       & 9--10 & 9.44 & 0.0058 & 0.0155 & 0.0010 \\
         &       & 10--12 & 10.8 & 0.0208 & 0.0181 & 0.0016 \\
         &       & 12--20 & 13.6 & -0.0099 & 0.0286 & 0.0011 \\
                \\
$\eta$   &     \ptau & 2--3 & 2.64 & 0.01279 & 0.00665 & 0.00306 \\
         &       & 3--4 & 3.44 & -0.00255 & 0.00377 & 0.00115 \\
         &       & 4--5 & 4.41 & -0.00168 & 0.00448 & 0.00117 \\
         &       & 5--6 & 5.4 & -0.00810 & 0.00667 & 0.00183 \\
         &       & 6--7 & 6.41 & 0.0064 & 0.0108 & 0.0029 \\
         &       & 7--8 & 7.42 & -0.0056 & 0.0170 & 0.0026 \\
         &       & 8--10 & 8.74 & -0.0122 & 0.0216 & 0.0002 \\
         &       & 10--20 & 11.7 & 0.0615 & 0.0351 & 0.0021 \\
                \\
$\eta$   &       \ptal & 2--3 & 2.64 & -0.0044 & 0.0107 & 0.0046 \\
         &       & 3--4 & 3.46 & 0.00043 & 0.00575 & 0.00219 \\
         &       & 4--5 & 4.42 & -0.00278 & 0.00686 & 0.00050 \\
         &       & 5--6 & 5.41 & -0.0022 & 0.0104 & 0.0031 \\
         &       & 6--7 & 6.42 & -0.0032 & 0.0163 & 0.0055 \\
         &       & 7--8 & 7.42 & 0.0468 & 0.0260 & 0.0004 \\
         &       & 8--10 & 8.74 & 0.0017 & 0.0318 & 0.0027 \\
         &       & 10--20 & 11.7 & 0.0395 & 0.0464 & 0.0133 \\  
            \end{tabular}  \end{ruledtabular}
        \label{tab:pA_finalAN}
\end{table*}

Similar to the recent $p$$+$$p$ $\pi^{0}$ and $\eta$ TSSA 
analysis~\cite{PPG234}, the transverse single-spin asymmetry $A_N$ is 
determined with the ``relative-luminosity'' formula, which is calculated 
separately for the two detector arms.  This yields measurements from two 
independent data sets that are verified for consistency and then averaged 
to obtain the final result. The equation for the relative-luminosity 
TSSA is
\begin{equation}
A_N = \frac{1}{P\left<cos(\phi)\right>} \frac{{N^{\uparrow}}-\mathcal{R}{N^{\downarrow}} }{{N^{\uparrow}}+\mathcal{R}{N^{\downarrow}}},
\label{eq:relan}
\end{equation}
where $P$ is the beam polarization, and $\mathcal{R}$ is the relative 
luminosity, defined as the ratio of integrated luminosities between 
the bunches with $\uparrow$ and $\downarrow$ spin states and measured 
by the BBC detectors. $\left<cos(\phi)\right>$ is the acceptance 
factor which reflects the detector azimuthal coverage, calculated 
separately for each $p_{T}$ bin and spectrometer arm, and $N$ refers 
to the yields, with the arrows referring to the up ($\uparrow$) or 
down ($\downarrow$) polarization of the proton beam.

Another method to calculate the asymmetry is the ``square-root'' 
formula, which is used as a cross check. The square-root formula is 
defined as
\begin{equation}
A_N = \frac{1}{P\left<cos(\phi)\right>} \frac{\sqrt{N_L^{\uparrow}N_R^{\downarrow}}-\sqrt{N_L^{\downarrow}N_R^{\uparrow}} }{\sqrt{N_L^{\uparrow}N_R^{\downarrow}}+\sqrt{N_L^{\downarrow}N_R^{\uparrow}}},
\label{eq:sqrtan}
\end{equation}
and it is used to calculate the asymmetry for both spectrometer arms 
simultaneously, where the $L$ and $R$ subscripts of $N$ correspond to 
yields measured in the left and right detector arms respectively, with 
respect to the polarized-proton-going direction. This results in only 
one measurement of $A_{N}$ that can be compared with the weighted 
average of the left and right relative-luminosity asymmetry calculation. 
The comparison of results using \eq{eq:relan} and \eq{eq:sqrtan} was 
used as a cross check, with corresponding systematic uncertainties 
discussed below.

As an additional cross check, the TSSA is calculated as a function of 
$\phi$, in which case a cosine modulation is fit to extract the 
asymmetry. This was found to be statistically consistent with the main 
asymmetry results.

The measured asymmetries were corrected for the background as follows, 
\begin{equation}
A_N^{\rm sig} = \frac{A_N-r \cdot A_N^{\rm BG}}{1-r},
\label{eq:sb}
\end{equation}
where $r$ is the background fraction under the \piz or $\eta$ peaks, 
calculated from the background fits ([green] solid lines) shown in 
\fig{fig:invmass}. $A_N^{\rm BG}$ is the background asymmetry, which was 
evaluated from the side bands on both sides of the \piz and $\eta$ 
peaks, also shown in \fig{fig:invmass}. The background asymmetry was 
consistent with zero in all \pt bins and all collision systems for both 
the \piz and $\eta$ mesons.

Asymmetries were calculated separately for each accelerator fill, during 
which the detector performance and beam conditions are considered to be 
relatively stable. The final asymmetry was obtained from the weighted 
average over the accelerator fills.

The possible sources of systematic uncertainties considered are 1) the 
background contribution $r$ in \eq{eq:sb}, and 2) possible variation in 
detector performance and beam conditions. A systematic uncertainty on 
the background fraction is quantified by varying the fit ranges used to 
calculate $r$, and computing how much the background-corrected asymmetry 
changes. The variations in detector performance and beam conditions, 
including uncertainty on the relative luminosity, were tested by 
comparing results calculated with the ``relative-luminosity formula" 
\eq{eq:relan} and the ``square-root formula" \eq{eq:sqrtan}, and with a 
technique known as ``bunch shuffling''~\cite{bunchShuffling}. The 
asymmetries calculated with the different formulas were found to be 
statistically consistent when taking into account the correlation 
between datasets. However, a conservative systematic uncertainty 
calculated as the absolute value of the difference in central values is 
assigned. In the bunch-shuffling procedure, the polarization of each 
bunch is randomly assigned to be up or down, and a distribution of $A_N$ 
in each $p_{T}$ bin is obtained by repeating the procedure 10,000 
times~\cite{bunchShuffling}. While most $p_{T}$ bins were found to be 
consistent with the statistical variation, some of the lower $p_{T}$ 
bins for both the $\pi^{0}$ and the $\eta$ in both \pA collision systems 
included up to 15\% variation beyond what was expected from statistical 
fluctuations. To account for this effect, an additional systematic 
uncertainty was assigned in any $p_{T}$ bin showing variations 
significantly beyond expected statistical fluctuations.

\section{Results and discussion}

\figure{fig:ANpT} shows the measurement of $A_N$ for \piz and $\eta$ 
mesons in \ptau and \ptal collisions at \sqsn = 200 GeV. The measured 
asymmetries in \ptau and \ptal are consistent with zero across the 
entire \pt range for both the \piz and $\eta$ mesons. 
Table~\ref{tab:pA_finalAN} lists the asymmetries, statistical 
uncertainties, and total systematic uncertainties for the $\pi^{0}$ and 
$\eta$ mesons in \ptau and \ptal collisions.

The TSSA measurements presented here probe the complex dynamics of 
partons within a nucleus. Measurements of asymmetries with heavy nuclei 
have not been performed at collider energies before 2015. 
Therefore, it is unclear to what extent the nuclear environment affects 
TSSAs. Collisions with a nucleus explore spin-momentum correlations in 
an environment where larger multiplicities and stronger color fields 
could play an additional role. In a factorized picture, initial-state 
spin-momentum correlations in the polarized proton cannot be affected by 
the presence of a nucleus; however, it is possible for final-state 
spin-momentum correlations in the process of hadronization to be 
affected as the scattered parton passes through the nuclear matter. 
Allowing for factorization-breaking effects, the larger color field of 
the nuclear remnant in \pta collisions as compared to the proton remnant 
in \ptp collisions could potentially modify the observed 
asymmetries~\cite{fact_break_2010,extra_asymmetries}. Neutral-pion 
measurements in the forward region~\cite{STAR:2020grs} and 
charged-hadron measurements in the intermediate rapidity 
region~\cite{PPG215, PPG226} show sizable TSSAs in \ptp collisions, with 
moderate nuclear modifications in \pta for the former and 
strong nuclear modifications in \pta for the latter. In 
contrast, the $\pi^0$ and $\eta$ meson asymmetries at midrapidity are 
consistent with zero in all collision systems, showing no difference 
between \ptp and \pta collisions. \par

\section{Summary}

The data presented here were motivated by the outstanding questions 
regarding the physical origin of transverse single-spin asymmetries. The 
TSSAs of midrapidity \piz and $\eta$ mesons were measured in \ptau, and 
\ptal collisions at \sqsn $=200$ GeV by the PHENIX experiment at RHIC.  
The measured asymmetries are consistent with zero up to very high 
precision in both collision systems for both meson species. The data 
presented here will contribute to the understanding of transverse spin 
phenomena in the more complex environment present in proton-nucleus 
collisions. In particular, we find that at midrapidity the presence of a 
heavy nucleus in the collision does not significantly modify the 
magnitude of the measured TSSAs.

%%%%%%%%%%%%%%%%%%%%%%%%%  Acknowledgements 

% \section*{ACKNOWLEDGMENTS}   % <-- uncomment for PRC or PRD only

%%%%%%%%%%%%%%%%%%%%%%  ACKNOWLEDGMENTS}  %%%%% MGS22a version
%% 2018 change in Korea
%%% 2021 change in dropping Brazil, Germany, and Pakistan, because
%%%      they no longer have active MGS and left PHENIX before 2015

\vspace{3.0cm}

\begin{acknowledgments}

We thank the staff of the Collider-Accelerator and Physics
Departments at Brookhaven National Laboratory and the staff of
the other PHENIX participating institutions for their vital
contributions.  
We acknowledge support from the Office of Nuclear Physics in the
Office of Science of the Department of Energy,
the National Science Foundation,
Abilene Christian University Research Council,
Research Foundation of SUNY, and
Dean of the College of Arts and Sciences, Vanderbilt University
(U.S.A),
Ministry of Education, Culture, Sports, Science, and Technology
and the Japan Society for the Promotion of Science (Japan),
Natural Science Foundation of China (People's Republic of China),
Croatian Science Foundation and
Ministry of Science and Education (Croatia),
Ministry of Education, Youth and Sports (Czech Republic),
Centre National de la Recherche Scientifique, Commissariat
{\`a} l'{\'E}nergie Atomique, and Institut National de Physique
Nucl{\'e}aire et de Physique des Particules (France),
J. Bolyai Research Scholarship, EFOP, the New National Excellence
Program ({\'U}NKP), NKFIH, and OTKA (Hungary),
Department of Atomic Energy and Department of Science and Technology
(India),
Israel Science Foundation (Israel),
Basic Science Research and SRC(CENuM) Programs through NRF
funded by the Ministry of Education and the Ministry of
Science and ICT (Korea).
Ministry of Education and Science, Russian Academy of Sciences,
Federal Agency of Atomic Energy (Russia),
VR and Wallenberg Foundation (Sweden),
University of Zambia, the Government of the Republic of Zambia (Zambia),
the U.S. Civilian Research and Development Foundation for the
Independent States of the Former Soviet Union,
the Hungarian American Enterprise Scholarship Fund,
the US-Hungarian Fulbright Foundation,
and the US-Israel Binational Science Foundation.

\end{acknowledgments}

%\clearpage

%%%%%%%%%%%%%%%%%%%%%%%%%%%  References 

%\bibliography{ppg204x1}   

%%% =============== end of main document  ========================

 
\end{document}